\documentclass[
    ,final            
  ]
  {aipproc}

\layoutstyle{6x9}

\begin{document}

\title{Four remarks on spin coherent states}

\author{Anna Baecklund}{
address={Teoretisk fysik, Kungliga Tekniska H\"ogskolan, 106 91 Stockholm, Sweden}
}

\author{Ingemar Bengtsson}{
  address={Fysikum, Stockholms Universitet, 106 91 Stockholm, Sweden}
}


\begin{abstract}
We discuss how to recognize the constellations seen in the 
Majorana representation of quantum states. Then we give explicit 
formulas for the metric and symplectic form on $SU(2)$ orbits 
containing general number states. Their metric and symplectic areas 
differ unless the states are coherent. Finally we discuss some 
patterns that arise from the Lieb-Solovej map, and for dimensions 
up to nine we find the location 
of those states that maximize the Wehrl-Lieb entropy. 
\end{abstract}

\maketitle


\section{Introduction}

Much of quantum mechanics concerns the action of some group (perhaps 
under experimental control) on Hilbert space. The group $SU(2)$ provides a 
simple and instructive case. We will make four remarks that we believe are 
new, and worth making. They are detailed in the abstract. 

To begin we fix a definite representation of 
$SU(2)$ by means of Schwinger's oscillator representation \cite{Schwinger}. 
It starts with two commuting pairs of creation and annihilation operators 

\begin{equation} [a_+,\bar{a}_+ ] = [a_-,\bar{a}_- ] = 1 \ . \end{equation}

\noindent There are orthonormal basis states 

\begin{equation} |n_+,n_-\rangle = \frac{(\bar{a}_+)^{n_+}}{\sqrt{n_+!}}\frac{(\bar{a}_-)^{n_-}}{\sqrt{n_-!}}|0,0\rangle 
\ . \label{factorials} \end{equation}

\noindent We refer to these states as number states. In terms of the 
oscillators we can write the $SU(2)$ Lie algebra generators as 
well as a number operator $\hat{n}$: 

\begin{equation} S_x = \frac{1}{2}(\bar{a}_+ a_- + \bar{a}_- a_+) \hspace{5mm} 
S_y = \frac{1}{2i}(\bar{a}_+ a_- - \bar{a}_- a_+) \hspace{5mm} 
S_z = \frac{1}{2}(\bar{a}_+ a_+ - \bar{a}_- a_-) \end{equation}

\begin{equation} \hat{n} = \bar{a}_+ a_+ + \bar{a}_- a_- \ . 
\end{equation}

\noindent The Hilbert space is infinite dimensional but we restrict ourselves to 
eigenspaces of $\hat{n}$, that is we fix $n = n_++n_-$ and obtain an irreducible representation of dimension $N = n+1$. If the physics 
is that of a spin system we set $n_+ = j+m$ and $n_- = j-m$. 

An alternative way of seeing how $SU(2)$ acts on ${\bf C}^{n+1}$ is to observe that 
the components of the vectors are in one-to-one correspondence with the coefficients 
of an $n$th order polynomial in an auxiliary complex variable $\zeta$ \cite{Majorana, 
Penrose, Bacry}. Up to an irrelevant complex factor such polynomials are determined 
by their $n$ complex roots, hence by $n$ possibly coinciding points in the complex plane 
taken in any order. Finally stereographic projection turns these $n$ points into $n$ 
unordered points on a sphere. If a root sits at $z_i$ the latitude and longitude of the 
corresponding point on the sphere are given by 

\begin{equation} z_i = \tan{\frac{\theta_i}{2}}e^{i\phi_i} \ . \end{equation} 

\noindent The South Pole is at $z = \infty$. If the polynomial has $k$ roots there 
its degree is only $n-k$. A collection of $n$ unordered points on the sphere is 
called a constellation of stars, since---in Penrose's original application---the 
sphere was literally to be identified with the celestial sphere \cite{Penrose}. 

The charming simplicity of the idea is compromised just a little by the care needed 
to ensure that an $SU(2)$ transformation in Hilbert space corresponds to a rotation 
of their celestial sphere. A state is described interchangeably as a vector $|\psi\rangle$ 
or as a polynomial $w(\zeta )$ in an auxiliary variable $\zeta$ through 

\begin{equation} |\psi \rangle = \sum_{k=0}^n \psi_k |n-k,k\rangle \hspace{5mm} 
\leftrightarrow \hspace{5mm} 
w(\zeta ) = \sum_{k=0}^n(-1)^k\sqrt{{n \choose k}}\psi_k\zeta^{n-k} \ . 
\end{equation}

\noindent Up to an irrelevant factor the polynomial admits the 
unique factorization 

\begin{equation} w(\zeta ) = (\zeta -z_1)(\zeta - z_2)\cdot \dots \cdot (\zeta -z_n) = 
\sum_{k=0}^n(-1)^ks_k(z)\zeta^{n-k} \ . 
\end{equation}

\noindent Here $s_k(z)$ is the $k$th symmetric function of the $n$ roots $z_i$. 
These conventions answer all our needs. Given a constellation of $n$ stars on the 
sphere we can reconstruct the vector up to an irrelevant constant in terms of 
symmetric functions of the roots. To go the other way we must solve an $n$th order 
polynomial equation. 

Spin coherent states are those for which all the stars coincide,

\begin{equation} w(\zeta ) = (\zeta - z)^n \hspace{5mm} \leftrightarrow \hspace{5mm} 
|z \rangle = \frac{1}{(1+|z |^2)^{\frac{n}{2}}} \sum_{k=0}^n 
\sqrt{{n \choose k}}z^k|n-k,k\rangle \ . \end{equation}

\noindent Here we took care to normalize the states. An arbitrary normalized state is 

\begin{equation} |\psi\rangle = N \sum_{r=0}^\infty \frac{s_k(z_1,\dots ,z_n)}{\sqrt{{n \choose k}}}
|n-k,k\rangle \ , \label{Norm} \end{equation}

\noindent where $N = N(z_1,\dots , z_n)$ is a normalizing factor to be 
worked out. We then define the everywhere non-negative Husimi function 

\begin{equation} Q_\psi (z,\bar{z}) = \langle \psi|z\rangle \langle z|\psi \rangle 
\hspace{5mm} \Rightarrow \hspace{5mm} \frac{n+1}{4\pi}\int d\Omega \ Q_\psi = 1 \ , \end{equation}

\noindent where $d\Omega$ is the measure on the unit sphere and the state is normalized. 
The implication happens because the coherent states form a POVM.  
Hence, for all choices of the state the Husimi function is a probability 
distribution on a sphere---namely on the set of coherent states. Its zeroes occur 
antipodally to the roots of the polynomial $w(\zeta )$ that defines the state, at 

\begin{equation} \omega_i = - \frac{1}{\bar{z}_i} \ . \end{equation}

\noindent Finally the Wehrl entropy of a pure quantum state is defined as 

\begin{equation} S_W(\psi ) = - \frac{n+1}{4\pi}\int d\Omega \ Q_\psi 
\ln{Q_\psi} \ . \end{equation}

\noindent Lieb conjectured \cite{Lieb} and Lieb and Solovej proved \cite{Solovej} (after an 
interval of 35 years during which many people thought about it) that this Wehrl entropy 
attains its minimum at the spin coherent states, thus singling out the latter as ``classical''.

This was brief. Details can be found in books \cite{BZ}, and elsewhere. 

\section{First remark: star gazing} 

Given a constellation of stars, can we recognize the corresponding quantum state 
without performing a calculation? Sometimes yes. We recognize the number states, we 
can sometimes see 
at a glance whether two states are orthogonal, and we can always recognize 
the time reversed state \cite{twistors}. 

To these cases we add constellations of type $k_{\rm N}$--$k_1$--$k_{\rm S}$, 
$k_{\rm N}$--$k_1$--$k_2$--$k_{\rm S}$, 
and so on, meaning that we place $k_{\rm N}$ stars at the North Pole, $k_i$ stars 
on regular polygons at some fixed latitudes, and $k_{\rm S}$ stars at the South 
Pole. (The vertices of the Platonic solids provide examples.) Let $q$ be a primitive 
$k_1$th root of unity. The configuration $k_{\rm N}$--$k_1$--$k_{\rm S}$ gives the 
polynomial  

\begin{equation} \zeta^{k_{\rm N}}(\zeta -\omega_1)(\zeta -\omega_1 q)\cdot \dots 
(\zeta- \omega_1q^{k_1-1}) = \zeta^{k_{\rm N}}\left( \zeta^{k_1} + (-)^{k_1}\omega_1^{k_1}
\right) \ . \end{equation}

\noindent The equality holds because all but two of the symmetric 
functions in $q^r$ vanish. The resulting (unnormalized) vector is 

\begin{equation} |\psi \rangle = \sqrt{(k_{\rm N}+k_1)!k_{\rm S}!}
|k_{\rm N}+k_1,k_{\rm S}\rangle + \omega_1^{k_1}\sqrt{k_{\rm N}!(k_1 + k_{\rm S})!}
|k_{\rm N}, k_1 + k_{\rm S}\rangle \ . \end{equation}

\noindent In fact, by varying the latitude and rotating the polygon we sweep out 
the entire two dimensional subspace spanned by the two number states.

From the configuration $k_{\rm N}$--$k_1$--$k_2$--$k_{\rm S}$ we obtain a 
four parameter family of states in a subspace spanned by four number states. 
If $k_2 = k_1$ two of the number states coincide but there are still four 
free parameters, and we 
sweep out an entire subspace spanned by only three number states.

\section{Second remark: orbits of number states}

The Majorana representation is ideally suited to study orbits under $SU(2)$ 
\cite{Bacry}. To find the orbit to which a given constellation belongs, just perform 
an arbitrary rotation of the celestial sphere. The set of constellations that appear 
in this way is the orbit. Since the group is three dimensional, so is a typical orbit. 
Number states, where all the stars sit at an antipodal pair of points, are exceptional 
and form two dimensional orbits. Intrinsically they are spheres, with antipodal points 
identified if $k = n/2$. 

Now recall that in projective Hilbert space (equipped with the Fubini-Study metric) 
a two dimensional subspace is intrinsically a Bloch sphere, of radius $1/2$, and also 
a K\"ahler manifold. The orbit of coherent states is a K\"ahler manifold too, but of 
a different radius. They form a rational curve in projective space \cite{Brody}, 
while the subspaces form projective lines. What about the orbits 
containing general number states? Since they are isolated orbits under the isometry 
group it immediately follows from a theorem in differential geometry \cite{Hsiang} 
that they are minimal submanifolds of projective space. To work out their intrinsic 
geometry we place $n-k$ stars at the point $z_1$ and $k$ stars at the antipode, and 
calculate

\begin{equation} \psi_1(z) = \langle \psi_1|z\rangle = \sqrt{{n \choose k}}|z_1|^{n-k}
\frac{(z-z_1)^k(z + 1/\bar{z}_1)^{n-k}}{(1+|z_1|^2)^{\frac{n}{2}}(1+|z|^2)^{\frac{n}{2}}} 
\ . \end{equation}

\noindent Next we take three such states and evaluate the Bargmann invariant 

\begin{equation} \langle \psi_0|\psi_1\rangle \langle \psi_1|\psi_2\rangle \langle 
\psi_2|\psi_0\rangle = \cos{D(0,1)}\cos{D(1,2)}\cos{D(2,0)}
e^{iA_{\Delta}} \end{equation}

\noindent to second order in the position of the stars. Here $D$ denotes the length 
of the geodetic edges of the triangle whose symplectic area is $A_\Delta$. 
Thus we obtain the intrinsic metric and the symplectic form on the orbit, 

\begin{equation} ds^2 = \frac{n+2k(n-k)}{4}\left[ \frac{4dz \otimes_{\rm sym} 
d\bar{z}}{(1+|z |^2)^2}\right] \ , 
\hspace{8mm} \Omega = \frac{n-2k}{4}\frac{i4dz \wedge d\bar{z}}{(1+|z |^2)^2} 
 \ . \end{equation}
 
\noindent The metric on the unit sphere appears between square brackets. Metrically 
the area of the orbit grows as $k$ approaches $n/2$, but 
symplectically it shrinks. When $k = 0$ the states are coherent, the two areas agree 
and we have a K\"ahler metric on the coherent state orbit. For even $n$ and $k = n/2$ 
the symplectic form vanishes. In this sense---which makes more sense than it seems 
to at first sight---the coherent state orbit contains the ``classical'' states 
\cite{Klyachko, Barnum, Kus}.     

\section{Third remark: the Lieb-Solovej map}

In their proof of the Lieb conjecture \cite{Lieb} Lieb and Solovej \cite{Solovej} 
introduced a completely positive map that, in a sense, allows us to 
approach the classical limit in stages. It maps density matrices acting on 
${\cal H}^{n+1}$ to density matrices acting on ${\cal H}^{m+1}$. If $m = n+1$ it is  

\begin{equation} \Phi^1(\rho ) = \frac{1}{n+2}\left( a_+^\dagger \rho a_+ + a_-^\dagger \rho a_- 
\right) \ , \end{equation}

\noindent and $\Phi^{m-n}$ is defined by iteration if $m > n+1$. It is easy to see that 
this map is trace preserving, and one also proves the key fact that 

\begin{equation} \Phi^{m-n}([S_i,\rho]) = [S_i,\Phi^{m-n}(\rho)] \ . \end{equation}

\noindent Therefore states on the same $SU(2)$ orbit in ${\cal H}^{n+1}$ will map 
to states on the same $SU(2)$ orbit in ${\cal H}^{m+1}$, and the resulting 
density matrices are isospectral. If $\rho$ is a pure coherent state the eigenvalues 
of $\Phi^{m-n} (|z\rangle \langle z|)$ can be computed. The proof then hinges on 
the beautiful theorem saying that the resulting spectrum majorizes all other occurring 
spectra. A modest illustration of this remarkable result seems called for.

\begin{figure}[ht]
  \resizebox{15pc}{!}{\includegraphics{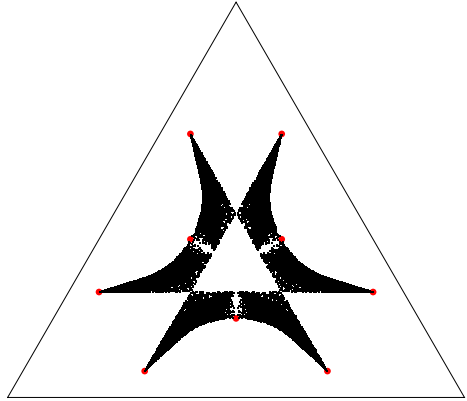}}
\caption{The spectra of 5 000 states under $\Phi^2$, $n = 3$.}
\end{figure}

\begin{figure}[ht]
  \resizebox{15pc}{!}{\includegraphics{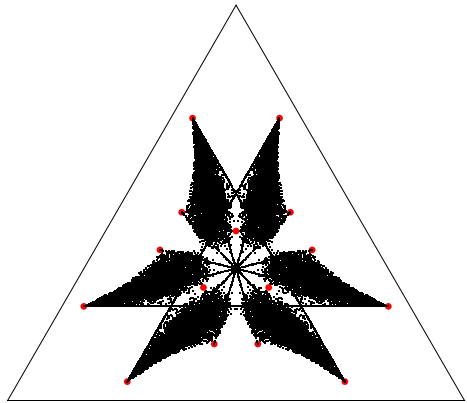}}
\caption{The spectra of 5 000 states under $\Phi^2$, $n = 4$.}
\end{figure}

Each time we apply the map $\Phi^1$ the rank of the density matrix goes up one step, 
so $\Phi^2$ applied to a pure state gives a spectrum described by a 
point in a two-dimensional simplex. The figures show results for 5 000 pure 
states chosen at random according to the Fubini-Study measure, for the initial 
dimensions 4 and 5. The straight lines that have been added are the spectra arising 
from linear combinations of two number states with $|n,k\rangle $ and $|n,k'\rangle$, 
$|k-k'| > 2$; for $n = 4$ the image can end up in the centre, but it is highly unlikely 
to do so. Images of the number states are marked by red dots, with the six coherent 
state images outermost. The latter do majorize all other spectra since all others  
lie in their convex hull, typically with a large 
margin.  
Interesting patterns arise in higher dimensions, but we have not done a systematic study.

\section{Fourth remark: Maximizing the Wehrl-Lieb entropy}

Once it is known that the Wehrl entropy attains its minimum at the 
coherent states it is irresistible to ask where it attains its maximum. In the Majorana 
representation the problem is to choose a constellation of points on the sphere that 
maximizes a particular function. Simpler, but still very difficult, 
relatives of this problem include the Thomson problem of minimizing the electrostatic 
potential of $n$ electrons on a sphere and the Tammes problem of maximizing the minimal 
chordal distance between the electrons. The possible physical 
motivation behind our problem is shared by the Queens of Quantum problem to maximize 
the distance between a pure state and the convex hull of the coherent states \cite{Giraud}. 

We have not attempted a full scale optimization of the Wehrl entropy. The reason should 
be evident to anyone who has seen this function written out explicitly as a function 
of the positions of the Majorana stars \cite{Lee, Schupp}. Instead we have 
taken reasonable candidates for the maximum, including those known to solve the other 
problems, and then we have checked whether a local maximum of the Wehrl entropy 
results. In those cases where natural parameters appear in the constellations we have 
maximized over these parameters. For instance, when $n = 8$ the maximum of the Thomson 
and the Tammes problem is given by two squares on two distinct latitude circles. This is a 
configuration of type $0$--$4$--$4$--$0$, and by the first remark it lies in the 
subspace spanned by $|8,0\rangle$, $|4,4\rangle$, and $|0,8\rangle$. We can rotate the 
squares relative to each other and 
change the difference in latitude. The Thomson, Tammes and Wehrl problems are all solved 
by this configuration, but the latitudes differ. In this way we have 
convinced ourselves that the results in the accompanying table are correct. Full details 
can be found elsewhere \cite{Anna}. 

\

\begin{tabular}{|c||c|c|c|c|} \hline
Number & Maximum & Queen of & Thomson & Tammes \\ 
of stars & Wehrl & Quantum & & \\ \hline
3 & triangle & triangle & triangle & triangle  \\
4 & tetrahedron & tetrahedron & tetrahedron & tetrahedron \\ 
5 & $1$--$4$--$0$ & $1$--$4$--$0$ & $1$--$4$--$0$ & $1$--$4$--$0$ \\
6 & octahedron & octahedron & octahedron & octahedron \\
7 & $1$--$5$--$1$ & $1$--$5$--$1$ & $1$--$5$--$1$ & $1$--$3$--$3$--$0$ \\
8 & $0$--$4$--$4$--$0$ & looks odd & $0$--$4$--$4$--$0$ & $0$--$4$--$4$--$0$ \\ 
9 & $0$--$3$--$3$--$3$--$0$ & $0$--$3$--$3$--$3$--$0$  & $0$--$3$--$3$--$3$--$0$  & 
$0$--$3$--$3$--$3$--$0$ \\ \hline
\end{tabular}

\

\

We would like to compare the Wehrl and Queens of Quantum problem in Hilbert spaces of 
dimensions as high as those number of stars for which the Thomson problem has been 
studied \cite{Erber} However, in a note on spherical harmonics 
which is quite relevant here, Sylvester apologizes for treating some things sketchily 
because he was ``very much pressed for time and within twenty-four hours of steaming 
back to Baltimore'' \cite{Sylvester}. The twenty-first century is no less pressing 
than was the nineteenth, and we have not been able to go further.   

\section{Summary}

The Majorana representation of quantum states \cite{Majorana}, in Hilbert 
spaces of dimension $N$, as consisting of constellations of $N-1$ unordered 
stars on a sphere \cite{Penrose}, is very 
useful in all contexts where the group $SU(2)$ plays a dominating role. To set 
the stage we gave  a brief discussion of how to see where we are in Hilbert space, 
given such a constellation. Next we gave explicit formulas for the metric and 
symplectic form on orbits through general angular momentum eigenstates. These 
orbits are interesting because they are symplectic but not K\"ahler, except for 
the coherent state orbit which 
goes through the highest weight states. A small observation on the Lieb-Solovej 
map followed. This map was introduced \cite{Solovej} in order to prove that 
coherent states minimize the Lieb-Wehrl entropy. Given that it 
is irresistible 
to ask for those states that maximize it. We presented results on this 
question for dimensions $N \leq 9$. It would be interesting to see results 
for higher dimensions.

\begin{theacknowledgments}
We thank Johan Br\"annlund and Kate Blanchfield 
for asssistance. One of us thanks Andrei Khrennikov and Ekaterina Axelsson for yet 
another useful V\"axj\"o meeting.
\end{theacknowledgments}



\bibliographystyle{aipproc}   

\bibliography{sample}

\IfFileExists{\jobname.bbl}{}
 {\typeout{}
  \typeout{******************************************}
  \typeout{** Please run "bibtex \jobname" to optain}
  \typeout{** the bibliography and then re-run LaTeX}
  \typeout{** twice to fix the references!}
  \typeout{******************************************}
  \typeout{}
 }

\end{document}